\documentstyle{article}

\def\ben{\begin{equation}}
\def\een{\end{equation}}
\def\bea{\begin{eqnarray}}
\def\eea{\end{eqnarray}}
\input amssym.def
\input amssym.tex
\begin{document}

\hfuzz=100pt
\title{Born-Infeld particles and Dirichlet p-branes}
\author{G. W. Gibbons
\\
D.A.M.T.P.,
\\ Cambridge University, 
\\ Silver Street,
\\ Cambridge CB3 9EW,
 \\ U.K.}
\maketitle

\begin{abstract}
Born-Infeld theory admits finite energy point particle solutions
with $\delta$-function sources,
BIons. 
I discuss their role in the theory of Dirichlet $p$-branes
as the ends of strings intersecting the brane 
when the effects of gravity are ignored. 
There are also topologically non-trivial electrically neutral
catenoidal solutions looking like two
$p$-branes joined by a throat. The general solution
is a non-singular deformation of the catenoid
if the charge is not too large and a singular deformation of the
BIon solution for charges above that limit. 
The intermediate solution is BPS and Coulomb-like.
Performing a duality rotation we obtain monopole solutions, 
the BPS limit being a solution of the abelian Bogolmol'nyi equations.
The situation closely resembles that of sub and super extreme
black-brane solutions of the supergravity
theories. I also show that certain special Lagrangian
submanifolds of ${\Bbb C}^p$, $p=3,4,5$, may be regarded
as supersymmetric configurations
consisting of $p$-branes at angles joined by throats
which are the sources of global monopoles. Vortex solutions are also exhibited. 
\end{abstract}

\section{Introduction}

Recently a number of outstanding problems in physics
have been, and are currently being,
solved using techniques involving 
Dirichlet p-branes (see \cite{Bachasrev} for a brief review
and references). It is therefore desirable to
understand as much as one can of their basic properties.
In particular one is interested
in the how stringy and spacetime concepts are related.
At the classical level p-branes have been extensively studied
as solutions of the various super-gravity limits
of string/M-theory. In this paper I want to explore
a different limit in which the effects of gravity are
ignored but in which one still has 
non-trivial and hopefully physically relevant classical solutions. 
We shall discover that there is a rich set of solutions
whose properties resemble in a striking way some of those
of the the black p-brane solutions of supergravity theory.

Classically the equations of motion
are essentially a generalization of the minimal surface
equations 
with the added feature that there is an abelian, i.e. $U(1)$,
gauge field propagating
on the world volume. In what follows we shall consider the case
of very weak string coupling in which case one may ignore  the 
fields having the  p-brane as their source. Thus the D-p-brane  
moves in flat Minkowski spacetime with constant dilaton
and vanishing Kalb-Ramond 3-form .
 It is consistent to set this gauge
field to zero and indeed we then get the standard minimal submanifold
equations,
of which the simplest solution is a flat $p+1$-plane. 
It is physically  clear that this is the {\sl only} static
regular solution defined as a single valued graph over ${\Bbb R}^p$
which becomes planar at large spatial
distances. This is the content
of the various eeven stronger 
\lq Bernstein\rq  type theorems
in the mathematical literature on minimal 
submanifolds \cite{L,Osserman}.
In fact it is notorious that these
strong 
theorems only hold strictly if $p<7$ which
no doubt will ultimately turn out to have a stringy/M-theoretic explanation.
This does not mean to say that there are no non-trivial 
regular static solutions. There are. Of particular interest are
topologically non-trivial generalized catenoids consisting of  two parallel
asymptotically flat sheets
joined by a throat.

Perhaps surprizingly
the planar solution remains a solution when the gauge field 
no-longer vanishes. In that case the gauge field is governed by the
pure 
Born-Infeld action \cite{BI,B}. Again it is not difficult
to convince oneself that the only  regular
static source free
solution of Born-Infeld theory which falls off at large distances is the 
trivial
one.  As we shall see there are  also
 Bernstein type theorems on maximal hypersurfaces
in Minkowski spacetime which prove this rigorously \cite{Calabi,YC}.  However it is well known that Born-Infeld theory admits
finite energy static solutions
which were originally
proposed as classical models for the electron. 
These solutions are not everywhere source free like soliton
solutions but rather resemble what are sometimes called 
\lq elementary\rq solutions with pointlike sources like the fundamental string solution \cite{DH,DGHR}. Perhaps one might call them \lq BIons\rq.

The essential  point of Born's theory is that one distinguishes between 
the electric field ${\bf E}$ which is curl free and hence one may set
\ben
{\bf E} = -\nabla \phi
\een
and the electric induction ${\bf D}$
which satisfies\footnote{ In this paper we shall be using
units in 
which $2\pi \alpha ^\prime=1$.}
\ben
\nabla \cdot {\bf D}=4 \pi e \delta({\bf x}-{\bf a})
\een
where $\bf a$ is the location of the the source
and $e$ is its electric charge.
Obviously $\bf D$ blows up at the origin
but because, in the absence of magnetic fields,
 the electric field and the induction are related by
\ben
{\bf E}={ {\bf D} \over \sqrt { 1+{\bf D}^2} }
\een
the electric field tends to a finite value at the source.
This maximal field strength has a nice interpretation
 in string theory because
of a divergence in the rate of production of open strings by 
the Schwinger process\cite{BachasP}.

It is natural to interpret the sources of these
BIionic solutions as the ends of  electric flux-carrying strings
lying outside the brane in the way suggested  in \cite{stro}.
However a point of difference with that analysis is that
the strings described there do not so much end on branes but 
rather disappear
down their throats. This is appropriate in the 
the case that the curved geometry generated by the brane in the 
supergravity limit
is taken into account but as mentioned above in the limit
we are considering the gravitational field of the brane is ignored. 
In fact we will later describe solutions 
with non-vanishing scalars 
in which the strings look like  very thin tubes joining smoothly
onto the p-brane.

As explained in \cite{stro} there are constraints due
to charge conservation on what strings can end on what branes.
These constraints arise because the interaction
with the Kalb-Ramond 2-form is obtained by replacing 
the Maxwell 2-form $F_{\mu \nu}$ in the world volume action $S_p$
by ${\cal F} _{\mu \nu}= F_{\mu \nu}-B_{\mu \nu}$ where $B_{\mu \nu}$
is the pullback to the world volume of the Kalb-Ramond 2-form.
The electric charge  $e$ is given by a surface integral over any 
$p-1$ sphere
lying in the brane and surrounding the source
\ben
{ 1 \over  A _{p-1}} \int _{S^{p-1}}  \star_p D 
\een
where $A_{p-1}$ is the volume of a unit $p-1$ sphere and
$\star_p$ is the world volume Hodge duality
operator and the 2-form $D$ has components given by the 
variational derivative 
\ben
D^{\mu \nu} = - { \delta S_p  \over \delta F_{\mu \nu} }
\een
and is thus  the covariant form of the electric induction.

The fact that $S_{p-1}$ contains $F_{\mu \nu}$ and $B_{\mu \nu}$ only
in the combination ${\cal F} _{\mu \nu}= F_{\mu \nu}-B_{\mu \nu}$
means that
\ben
D_{\mu \nu}= J_{\mu \nu}
\een
where 
\ben
J^{\mu \nu} = - { \delta S_p \over \delta B_{\mu \nu}} 
\een 
is the distributional current 2-form with support on the world-volume
which acts as a source for the Kalb-Ramond 3-form $H$.
Thus only fundamental, or $F$-strings can end on  $R\otimes R$ p-branes.
However by $SL(2,{\Bbb Z})$ invariance, both fundamental 
and Dirichlet or $D$-strings can end on the self-dual 3-brane,
the former ending on electric charges and the latter
on magnetic charges. This is consistent with the duality invariance of the
Born-Infeld equations of motion \cite{GR1}\cite{GR2}\cite{GZ}.

\section {Lagrangians and Equations of motion}

A Dirichlet-p-brane moving in a flat $d+1$ dimensional Minkowski
 spacetime ${\Bbb E}^{d,1}$  with constant dilaton and zero axion 
field is described by the embedding functions $z^M(x^\mu)$
and vector potential $A_\nu(x^\mu)$ as a function of the $p+1$ dimensional
world-volume coordinates $x^\mu$. Greek  indices thus run 
from $0$ to $p$ while upper case Latin indices run from $0$ to $d$.
The equations of motion are obtained from the Dirac-Born-Infeld
action\cite{Dirac,BI}:
\ben
-\int d^{p+1}x \sqrt {- {\rm det} ( G_{\mu \nu} + F_{\mu \nu} )}
\een
where 
\ben
F_{\mu \nu}= \partial _\mu A _\nu - \partial _\nu A _\nu
\een
is the electromagnetic field strength and 
\ben
G_{\mu \nu} =\eta_{MN}  \partial _\mu z^M \partial _\nu z^N 
\een
is the pullback of the Minkowski metric to the world volume.

The action is invariant under arbitrary diffeomorphisms of the 
world volume. One popular way of fixing this freedom
is to adopt the so-called \lq static gauge\rq for which the 
world volume coordinates
are equated with the first $p+1$ spacetime coordinates, i.e. 
\ben
z^M=x^\mu, ~~~~~~M=0,1,\dots ,p.
\een
Calling  the remaining \lq transverse \rq  coordinates $y^m$, i.e setting
\ben
z^M=y^m, ~~~~~~M=p+1,\dots d,~~~~m=p+1,\dots d,
\een
the Dirac-Born-Infeld action becomes
\ben
-\int d^{p+1}x \sqrt {-{\rm det} ( \eta_{\mu \nu} + \partial _\mu y^m \partial _\nu y^m+ F_{\mu \nu} )}.
\een

When speaking of a Born-Infeld Lagrangian  $L$ it is convenient to  
take off an additive constant so that $L$ vanishes for zero scalar and vectorfields. Thus we define
\ben
L= 1-\sqrt {-{\rm det} ( \eta_{\mu \nu} + \partial _\mu y^m \partial _\nu y^m+ F_{\mu \nu} )}
\een
It is important to realize that imposing 
the static gauge, or writing the equations in what mathematicians call 
in this context non-parametric form, may not always be possible globally. It assumes that
the brane may be thought of as a single valued graph
over ${\Bbb R}^p$. In particular that it is topologically trivial.
It will turn out in the examples that we shall encounter 
that this assumption is not valid. This will be reflected in spurious
singularities in some of our solutions when expressed in static gauge.
Nevertheless, since it is a very convenient guage for calculations,
we will frequently adopt it.

One might have thought that the transverse coordinates $y^m$ and the vector
field $A_\mu$ are inextricably coupled in that if one is non-zero then the 
other cannot vanish but this is not true. For example
one gets a {\sl consistent} set of solutions   
by setting the transverse coordinates to zero $y^m=0$ 
and requiring that the vector field satisfy the Born-Infeld
equations in flat $p+1$-dimensional Minkowski spacetime obtained by varying the action
\ben
-\int d^{p+1}x \sqrt{ -{\rm det} ( \eta_{\mu \nu} + F_{\mu \nu} )}.
\een 
The reason that this is a consistent truncation is that
the action is invariant under each of the
$d-p$ reflections given by 
\ben
y^m\rightarrow -y^m.
\een
The equations of motion for $y^m$ therefore involve 
 no terms of order zero, i.e. sources, and are trivially solved
by $y^m=0$.

As similar symmetry argument, (valid in any choice of world volume coordinates) shows that  one may consistently set
the vector fields to zero and obtain the Dirac action
\ben
\int d^{p+1}x \sqrt {-{\rm det} (\eta _{MN} \partial_\mu  z^M \partial _\nu z^N } ).
\een

In this case the symmetry we need is the invariance of the determinant
\ben
{\rm det} ( G_{\mu \nu} + F_{\mu \nu} ) 
\een
under transposition
\ben
 ( G_{\mu \nu} + F_{\mu \nu} )\rightarrow  ( G_{\mu \nu} + F_{\mu \nu} )^t=
 ( G_{\mu \nu} - F_{\mu \nu} )
\een
which is therefore equivalent to invariance under
reversal of the Maxwell field:
\ben
F_{\mu \nu} \rightarrow -F_{\mu \nu}.
\een

In fact one might have seen this result from another surprizing
property of the Dirac-Born-Infeld action. That is, it may be obtained from the
pure Born-Infeld action in $d+1$ dimensions 
\ben
-\int d^{d+1} z \sqrt {-{\rm det} ( \eta _{MN} + F_{MN} )} 
\een
by assuming that the vector field $A_N(z^M)$ depends
only on $p+1$ coordinates. One needs the following
identity for  determinants 
\ben
\left | 
\pmatrix {
N&-A^t \cr A&M \cr } \right | = 
\left | M \right | \left | N+ A^t M^{-1} A \right |
= \left | N \right | \left |M+ A N^{-1} A^t \right |,
\een
where as before $^t$ denotes transposition.
If one takes $M = \eta_{mn} = \delta _{mn}$ 
and $A= \partial _\mu A_n$
one finds the action
\ben
-\int d^{p+1}x \sqrt {-{\rm det} ( \eta_{\mu \nu} + \partial _\mu A_m \partial _\nu A_m+ F_{\mu \nu} )}.
\een
 
As is well known  we may identify the $d-p$ transverse components of the 
vector $A_n$ with the $d-p$ transverse coordinates $y^m$ of the p-brane.
A row expansion of the original determinant reveals that
one may consistently set to zero any of the possible transverse
polarizations $A_m$ because for any given $\mu$ and $m$, $\partial_\mu A_m$
appears only quadratically in the determinant.

\section{Pure Born-Infeld solutions}

We have seen that setting the transverse coordinates to zero gives a 
{\sl consistent} truncation of the Dirac-Born-infeld Lagrangian. In this section we shall
discuss some of the solutions 
of this theory and their properties. It will become clear
as we proceed that the solutions with
no Born-Infeld vectors play a similar role 
to pure gravity solutions in supergravity theories coupled to 
differential forms and scalars while 
the pure Born-Infeld solutions
are similar
 to solutions with no gravity or scalars. 
However there is one significant difference: in the supergravity 
gravity context,
while it may in some circumstances be consistent to
truncate the scalars, it is never strictly consistent to ignore gravity
completely. Moreover just as in supergravity
one may use solution generating transformations
to pass from pure gravity solutions to solutions with non-vanishing
differential forms, so in Dirac-Born-Infeld theory there are also
solution generating transformations. These take one from
pure scalar solutions to solutions with both scalars and vectors
and they also take one from pure Born-Infeld solutions
to solutions with non-vanishing scalars and vectors.

Of course one's greatest   familarity
is with solutions in three spatial dimensions
but the discussion will not be restricted to that case.

\subsection{Born-Infeld brane-waves}

A striking application of the simple matrix 
identity is obtained by taking $p=1$.
From the point of view of the $d+1$ dimensional field theory
one is studying the scattering of two 
light beams in non-linear electrodynamics. By Lorentz-invariance
the two beams can be taken to be moving in the positive or negative
$x^1$ direction. One might have  expected them to scatter non-trivially
but one finds that this is not so. Because  we must take $F_{\mu \nu}=0$ 
the action becomes the standard Nambu-Goto action for a free string
in $d+1$ spacetime dimensions:
\ben
-\int d^2 x \sqrt {-{\rm det} ( \eta_{\mu \nu} + \partial _\mu y^m \partial _\nu y^m  )}.
\een
 
It is well known that  left and right movers 
decouple in string theory it follows that the two beams pass through
one another with at most a time delay.  This fact
seems to be behind 
some early observation \cite{Schr} in four spacetime dimensions
where a form of this result was obtained 
in the particular case of two circularly polarized
beams with a single frequency.  
Later  work in \cite{BC1,BC2,BC3}
noted this property for  what is described
as a scalar equation \lq of Born-Infeld type\rq but
but the authors did not seem to realize that
their single component equation and its multi-component
generalizations were {\sl precisely} the equations of Born-Infeld theory
in this setting.  

\subsection{Domain walls and the Born-Infeld equation}

In the case $p=d-1$ there is just one transverse coordinate $y$
and the  corresponding Lagrangian is
\ben
L= \sqrt {1-\partial _\mu y \partial ^\mu y }.\label{eq:BItype}
\een

The resulting partial differential equation  is sometimes 
called the Born-Infeld or Born-Infeld-type equation and has been extemsively discussed in the literature.
In the case $d=3$  it describes a timelike membrane 
in four dimensional Minkowski spacetime and as such has been studied by
Bordemann and Hoppe\cite{Bordemann/Hoppe}. 

One obvious physical interpretation  of (\ref{eq:BItype})
is that it describe  interfaces between two symmetric domains.
The Born-Infeld interpretation (for example in four
spacetime dimensions)  is that  all electromagnetic
fields are independent of 
the third spatial coordinate and the only non-
vanishing fields are
\ben
E_3= \partial _t y
\een
\ben
B_1 = \partial _2 y
\een
and
\ben
B_2= -\partial _1 y.
\een
In other words a time-dependent electric field in the 
$3$-direction induces a magnetic induction field in the 
perpendicular $1,2$  directions.

Bordemann and Hoppe\cite{Bordemann/Hoppe} pointed 
(in fact for $d=3$ but the observation has an obvious generalization)
that one could regard the 
world sheet of the $d-1$ brane as a  timelike level set 
of a real valued function
$u(z^A)$ on $d+1$-dimensional Minkowski spacetime ${\Bbb E}^{d,1}$
which satisfies the manifestly $E(d,1)$-invariant
equation
\ben
\partial _A \Bigl ( { \partial ^A u \over \sqrt 
{ \partial _B u \partial ^B u}} \Bigl )=0.
\een
This equation may be derived from the $E(d,1)$ invariant  Lagrangian
\ben
L= \sqrt {-\partial ^A u \partial _A u }. \label{eq:Hoppe}
\een
The function $y$ is then obtained implicitly from the relation
\ben
u(x^\mu,y)= {\rm constant}.
\een
In fact $u$ need not satisfy the equation everwhere, 
merely restricted to the level set which gives rise to the weaker
condition:
\ben
\partial _A u \partial ^A u \partial _B \partial^B u=
\partial ^Au \partial ^B u \partial _B \partial_A u \Bigr |_{u={\rm constant}}.
\label{eq:Hoppe/Bordemann}
\een
Solutions of this condition will be used later to construct
BIonic crystals. For the time being we give two simpler examples:
Clearly if
\ben
u=-t^2+x^2+y^2+z^2 \label{eq:tcone}
\een
then $u=0$ gives the light cone of the origin
which is a  solution of the Bordeman-Hoppe
equation in ${\Bbb E}^{3,1}$. Rather surprizingly
it is also true that if
\ben
u=t^2+x^2+y^2-z^2 \label{eq:zcone}
\een
then the quadratic cone $u=0$ is also a solution in ${\Bbb E}^{3,1}$. 
For completeness we note that if
\ben
u=x^2+x^2+y^2-z^2 -\tau^2 \label{eq:econe}
\een
then the quadratic cone $u=0$ is  a minimal
3-brane in 
Euclidean 4-space ${\Bbb E}^{4}$.

One may wonder whether there are functions $u$ which
solve both the Hoppe-Bordemann equation (\ref{eq:Hoppe}) and the ordinary
wave equation $\partial^A \partial _Au=0$. This question has been studied
by Graustein \cite{Grau} in  ${\Bbb C}^3$
who found {\sl all} simultaneous
solutions in 
of these two equations. Some of his solutions are valid 
in ${\Bbb E}^{2,1}$. They include one
which depends on two arbitrary functions of a lightlike
variable,and which may easily generalized
to  ${\Bbb E}^{d,1}$ to depend on $d+$ arbitary functions 
of a lightlike variable. Choosing coordinates appropriately
it may be given the form:
\ben
u=p(t-z)-xa(t-z)-yb(t-z) -\dots ={\rm constant}. \label{eq:interface}
\een
The solution (\ref{eq:interface}) represents  a {\sl timelike} domain wall moving
at the speed of light along the z-axis. The functions
$p,a,b,\dots $ may be arbitary.

Another of Graustein's solutions is
\ben
u= (z+t)^{-{2 \over 3}} (t^2-x^2-z^2)^{ 1 \over 6}.
\een
In fact one may embed this solution in a family of solutions: 
\ben
u=(t+z)^p(t^2-x^2-z^2)^q
\een
will solve the Bordemann-Hoppe equation if
\ben
(p+q)(p+4q)=0
\een
and the wave equation if
\ben
1+2p+2q=0.
\een

In four dimensions one may find analogous solutions:
\ben
u=(z+t)^p(-t^2 +x^2 +y^2 +z^2)^q\label{eq:t-like}
\een
will solve the Bordemann-Hoppe equation if
\ben
(p+q)(p+6q)=0
\een
and the wave equation if
\ben
p+q+1=0.
\een
Also 
\ben
u=(z+t)^p (t^2 +x^2 +y^2 -z^2)^q \label{eq:z-like}
\een
will solve the Bordemann-Hoppe equation if
\ben
(p+q)(p-2q)=0
\een
and the wave equation if
\ben
p+q+1=0.
\een
The cases $p+q=0$ are in effect translations of  (\ref{eq:tcone})
and (\ref{eq:zcone}).
The example (\ref{eq:z-like}) with $p=2$ and $q=1$ was noted
by Hoppe (\cite{Hoppe2}).

Equation (\ref{eq:Hoppe/Bordemann}) is striking because it has
a topological character: it is invariant
under arbitrary one-dimensional reparameterizations $u \rightarrow f(u)$.
Thus in solutions (\ref{eq:t-like}) and (\ref{eq:z-like})
the  level sets depend only on the ratio of $p$ to $q$.

The  Lagrangian (\ref{eq:Hoppe}) for an interface
in ${\Bbb E}^{d,1}$ may be regarded as the strong coupling
limit\footnote{Restoring units this means that $2\pi \alpha ^\prime \rightarrow \infty$, i.e. we take the infinite slope limit in string theory.} of the Born-Infeld type Lagrangian (\ref{eq:BItype})
for an interface in ${\Bbb E}^{d+1,1}$. This is presumably related to
the fact that the strong coupling limit of the standard 
Born-Infeld Lagrangian picks out a topological invariant: the Pfaffian
$\sqrt { {\rm det} F_{\mu \nu}}$.

\subsection{Electrostatic Solutions and Maximal Hypersurfaces}

The previous results on timelike (d-2)-branes in ${\Bbb E}^{d,1}$ have
 companions involving spacelike hypersurfaces.
If one starts from the pure Born-Infeld theory in ${\Bbb E}^{d,1}$
and assumes only an electrostatic field is present so that
\ben
A_M= (\phi, 0,\dots,0)
\een
then the action becomes: 
\ben 
\int d^{d}x \sqrt {1 -|\nabla \phi |^2 }.
\een
As long as $|\nabla \phi |<1$, which means that the electrostatic field ${\bf E} =  -\nabla \phi$
is less than the maximal value allowed in Born-Infeld theory,
then one may regard $t=\phi( {\bf x})$ as the height function 
of a spaclike maximal hypersurface in Minkowski spacetime. 
The surface tips over and touches the light cone  precisely when the maximal field strength is attained. 

We shall now attempt to capitalize on the facts
that  maximal hypersurfaces have received much attention in 
the general relativity and differential geometry literature
\cite{KM1,M1,EH,ER,YC,Calabi,Bartnik}
and Born-Infeld electrostatics was extensively
studied in the 1930's \cite{B,B-B}
to obtain some insight
into both topics by exploiting this connection. 
The first obvious point
is that one  may use the Poincar\'e symmetries of the maximal
hypersurface problem in $d+1$ diimensional Minkowski spacetime to 
generate some useful new solutions of the electrostatic problem.
Put another way, our observation allows us to extend the obvious invariance
of equation under the Euclidean group $E(d)$ of isometries
of d-dimensional euclidean space ${\Bbb E}^d$ to a non-obvious or 
\lq hidden\lq  invariance under the Poincar\'e group $E(d,1) \supset E(d)$
of isometries of Minkowski spacetime ${\Bbb E}^{d,1}$.
This is , of course, closely related to some observations of Bachas\cite {Bachas} about D-branes in string theory in a related but not identical context.

As an illustration consider 
a uniform electric field
\ben
\phi =-zE,~~~~~ E<1
\een
where the  field of constant magnitude $E$ is taken to lie along the $z$-direction. This corresponds to a spacelike hyperplane and may be obtained
from the even more trivial solution $\phi =0$ by means of a Lorentz boost
in the $t-z$ 2-plane with velocity 
\ben
v=E.
\een
As noted above the maximum field strength condition $E<1$
arises from the  maximum velocity $v<1$.

According to the Bernstein type theorems of Calabi 
(valid for $n=2,3,4$ )\cite{Calabi} and of Cheng and Yau (valid for all $n>1)$
 \cite{YC} this solution is the {\sl only} one for which $\phi$ is a everywhere
non-singular and single valued. 
A weak form of this result follows easily from 
rather elementary  uniqueness
theorem of Pryce \cite{Pryce2}. Given two solutions he constructs the vector
\ben
{\bf G} = (\phi _1 - \phi_2) ( {\bf D}_1 - {\bf D}_2\bigl )
\een
and finds that
\ben
\nabla \cdot {\bf G} = u
\een
with
\ben
u= ( {\bf E}_1- {\bf E}_2 ) \cdot 
( 
{\partial L \over \partial {\bf E}_1 }
-
{\partial L \over \partial {\bf E}_2 } ) = 
( 
{\bf E}_1- {\bf E}_2
 )\cdot( 
{\bf D}_1- {\bf D}_2 ).
\een
Since $L$ is a positive  
stricly convex function of $\bf E$ 
, i.e. $L^\prime >0, L^{\prime \prime} >0$, 
then $u$ will be non-negative and vanish
only when $ {\bf E}_1 ={\bf E}_2$. Integration
of this identity over ${\Bbb R}^3$ with ${\bf E}_1$ 
being a uniform electric field and ${\bf E}_2$ being some other 
field which approaches the 
uniform field suficiently rapidly so as to make the boundary terms
vanish yields the uniqueness result \footnote{In fact uniqueness 
arguments of this kind also hold for more general non-linear electrodynamic
theories providing the associated Lagrangian $L$ satisfies an appropriate
convexity propery}.

It follows from the Gauss-Coddazi equations that the induced metric
\ben
g_{ij}=\delta _{ij}- \partial _i \phi  \partial _j \phi
\een
of a maximal 
hypersurface in Minkowski spacetime has non-negative Ricci-scalar.
On the other hand if the potential $\phi$ satisfies
\ben
\partial \phi = O( { 1\over r^2})
\een
at infinity the induced metric $g_{ij}$ has zero ADM mass.
Thus, by the positive mass theorem, the solutions must 
necessarily have singularities. Similarly, because there can be no
non-flat metric with non positive Ricci scalar on the three torus $T^3$,
there can be no non-singular triply periodic solutions. 
There are, as we shall see interesting solutions of both 
types with singularities, 
the singularities corresponding to electric charges.

\subsection{Inclusion of a magnetic field}

If one includes a magnetic field $\bf B$ the 
Born-Infeld 
Lagrangian
\bea
&&L=1-\nonumber \\
&&\sqrt{ 1-{\bf E}^2 + {\bf B}^2 -({\bf E}\cdot {\bf B} )^2 } \nonumber \\
\eea 
must be varied subject to the constraint that
the magnetic induction is divergence free
\ben
\nabla \cdot{\bf B}=0.
\een
This leads to the equation of motion 
(in the time independent case) that the magnetic field
\ben
{\bf H} =-{ \partial L \over \partial {\bf B}}
\een
is curl-free:
\ben
\nabla \times {\bf H}=0.
\een
As usual it is convenient to introduce a Lagrange 
multiplier $\chi$ to enforce
the constraint and perform a Legendre transformation
to yield  an unconstrained variational principle in terms
of the scalar field $\chi$ . Physically the scalar field $\chi$
is the magnetostatic potential 
\ben
{\bf H}= -\nabla \chi.
\een
The upshot of the Legendre tansformation 
is that one must vary the manifestly $SO(2)$ invariant
functional
\ben
\int d^3x \sqrt {1-(\nabla \phi)^2 -(\nabla \chi)^2 -(\nabla \phi)^2
(\nabla \chi)^2 + (\nabla \phi \cdot \chi)^2 }.
\een
This result may  also be obtained without explicitly
introducing the magnetostatic potential. We  can Legendre transform the 
Lagrangian to give 
\ben
{\tilde H ({\bf E}, {\bf H})}= L+ {\bf B}\cdot {\bf H}
\een
and then use the result to give the constitutive relations
\ben
{\bf D} ={ \partial {\tilde H} \over \partial {\bf E}}, 
\een
\ben
{\bf B} =-{ \partial {\tilde H}   \over \partial {\bf H}}.
\een
In this way one finds the manifestly 
electric-magnetic duality invariant formula\cite{B-B}
\ben
{\tilde H} =1- \sqrt {1-{\bf E}^2 -{\bf H}^2 +({\bf H} \times {\bf E})^2 }.
\een

Now if we were considering a static 3-brane moving in ${\Bbb E}^5$ in static
gauge we would extremize
\ben
{\rm det }~( \delta _{ij} + \partial _i y^1 \partial _j y^1 + \partial _i y^2 \partial _jy^2)
= 1+(\nabla y^1)^2 +(\nabla y^2)^2 +(\nabla y^1)^2
(\nabla y^2)^2 - (\nabla y^1 \cdot \nabla y^2)^2.
\een
Therefore formally one may regard $\phi$ and $\chi$ as
two timelike coordinates in ${\Bbb E}^{3,2}$ and
one may then check that the variational 
principle is that for a spacelike $3$-brane
in static gauge. The electric-magnetic duality 
is therefore seen directly as a geometrical rotational symmetry in this setting.

\subsection{Harmonic solutions}

For the uniform electric field ${\bf E} =-\nabla  \phi$  is a Killing vector
of the euclidean group $E(3)$. The most general Killing field of $E(3)$
is a screw rotation which may be supposed with no loss of generality
to be about the $z$-axis. This motivates checking
that in fact this  gives a solution
of the form
\ben
\phi = -{2J_m } \arctan ({y \over x}) -Ez
\een
Geometrically this defines a helicoid lying in the 
timelike hyperplane spanned by $(t,x,y)$. It ceases to be spacelike
inside the cylinder given by 
\ben
x^2 + y^2 = ({2J_m })^2 
\een

The physical interpretation of this solution if $E=0$ is  
that it describes the electric field generated by a magnetic current $J_m$
along the z-axis. Since the pure Born-Infeld
theory is invariant under electric magnetic duality rotations
\cite{B-B,GR1,GR2,GZ}, it may be easier to imagine an 
electric current
flowing along the z-axis generating a magnetic field. If
one approaches too close to the z-axis the magnetic field ${\bf H}$ 
 exceeds the maximum
allowed value and the solution breaks down. 
Presumably an electric current $J_e$ cannot,
according to  Born-Infeld theory, be contained within a wire of 
radius less than $2J_e $.

The solution for $\phi$ just presented is 
a harmonic function on ${\Bbb E}^3$. Moreover
every level set in ${\Bbb E}^3$ is a minimal surface.
By a result originally due to  Hamel
\cite{Nitsche},it is unique, because the electric field is a simultaneous
solution in ${\Bbb E}^3$
of the highly over constrained system:
\ben
\nabla \times {\bf E}=0,
\een
\ben
\nabla \cdot {\bf E}=0,
\een
and
\ben
{\bf E} \cdot \nabla  |{\bf E}|=0.
\een

The harmonic solutions just given generalize in an 
obvious way to higher dimensions
and different signatures.The exhustive list
of solutions given by Graustein in ${\Bbb C}^3$
\cite{Grau} quoted  earlier
in connection with the Bordemann-Hoppe equation (\ref{eq:Hoppe}) 
show that there are certainly other possiblities.

\subsection{ The BIon solution}

To be specfic we restrict ourselves to three spatial dimensions. 
the generalization to higher dimensions being immediate.
The solution is $SO(3)$ invariant:
\ben
\phi = f(r)= \int ^\infty _r { dx \over \sqrt { e^2 + x^4}}.
\een

Near infinity
\ben
\phi =   { e \over r} + O( { 1\over r^5}).
\een
Near the origin
\ben
\phi = \Phi  -r +{ r^5 \over 10 e^2 }+ O(r^9)
\een
where
\ben
\Phi= \sqrt e \int ^\infty _0 { dx \over \sqrt { 1 + x^4}}
\een
is the electrostatic potential difference beween the origin and infinity.
The electric charge $e$ of the solution is given by
\ben
e= { 1 \over 4 \pi} \int {\bf D}\cdot d \sigma= - { 1\over 4\pi}
\int  ({ \nabla \phi \over \sqrt { 1 +|\nabla \phi |^2 }})\cdot d \sigma 
\een
where the integral is over any closed 2-cycle enclosing the origin.

From  we observe that the solution is not smooth at the origin. 
The maximal hypersurface becomes 
tangent to the light cone through that point. 
Point like singularities
of maximal hypersurfaces of this type have been analysed
rather extensively in\cite{E}.

As noted by  Pryce \cite{Pryce2}, the 
BIon is the unique solution with a single singularity and
a  fixed charge $e$ or a fixed potential $\Phi$. This contrasts with
the behaviour  of Coulomb type solutions of Yang-Mills theory
which exhibit bifurcation phenomena indicative of instabilities.
Physically it is clear that a  consistency condition 
for
a particle like interpretation that non-uniqueness
and bifurcation  phenomena be absent.  
It has been argued \cite{DLS} that since the total
energy
of a single Born-Infeld particle in ${\Bbb E}^{p,1}$
scales like
\ben
e^{ p\over p-1} \label{eq:scaling}
\een
and since
\ben
(e_1+e_2)^{ p\over p-1}>e_1^{ p\over p-1}+e_2^{ p\over p-1}
\een
then they should be unstable against fission. 
This might be true if they were smooth singularity free solutions with no sources. However they {\sl do}  have 
sources and the strength of these sources is governed by charge quantization
conditions which arise from the perspective of string theory because
the particles are the ends of strings as described in \cite{stro}.
This means that a single particle carrying the lowest
possible charge should be stable. A particle carrying a multiple
of the lowest charge could however be stable against fission
into particles with the lowest possible charge.
These observations  \cite{DLS} are 
indicative of the fact that
these Born-Infeld particles are not BPS and moreover that they repel 
one-another. We shall see this is indeed true rather explicitly later. 
We shall also be saying more about the scaling relation (\ref{eq:scaling}) when we discuss
the virial theorem.

\subsection{BIon in  a uniform electric field}

At least if $n \ne 2$ this solution seems to be a genuinely new
one. It is most simply obtained by using the hidden Poincar\'e invariance.
One may  check that in the case $n=2$ it coincides with
a solution found by Pryce\cite {Pryce1} using the Weirstrass representation
for maximal surfaces in ${\Bbb E}^{2,1}$.  
The idea is to boost a BIon of charge $e$ with velocity $v$. if $\phi= g(x,y,z)$ is the solution then is is given implicitly in terms of $f(r)$
by

\ben
{ g-\Phi + vz\over \sqrt {1-v^2}} = 
f\bigl ( \sqrt {x^2 +y^2 +{( z+vg-v\Phi )^2\over 1-v^2 }}\bigr )-\Phi . 
\een
At large distances
\ben
\phi \approx \Phi (1-\sqrt{1-v^2}) -vz + 
{e\sqrt{1-v^2} \over r(1-v^2 \cos ^2 \theta)^ { 1\over 2}} \
\een
while near the origin
\ben
\phi \approx \Phi -r+ { (1-v\cos \theta )^6r^5 \over 10 e^2 (1-v^2 )^3 }+ 
O(r^9) 
\een
Thus the background electric field  is given by $E=v$ and the total charge
of the new solution is $ e \over \sqrt {1-v^2}$.
It is clear that given any other solution with total charge $q$
we may always append
a uniform electric field by boosting it with velocity $E=v$ to arrrive at 
a new solution with total charge $e\over \sqrt{1-v^2}$. 

We will discuss later the force necessary to prevent the
 particle accelerating.

\subsection{Accelerating solutions}

It would be nice to have an analogues of the various solutions
representing uniformly accelerating black holes in
external electric fields. One could then
study the possibility, using instanton methods,
of the pair-creation of BIon anti-BIon pairs
by the Schwinger process just as one can do with black holes.
Unfortunately no such explicit solutions are available at 
present. However there the solutions
(\ref{eq:t-like}) and (\ref{eq:z-like}) may be relevant here
since they are the level sets are taken
are taken into themselves by a boost in the $z-t$
2-plane.

\subsection{BIonic Crystallography}

Hoppe\cite{Hoppe1} has given a construction for  quadruply periodic maximal surfaces
in four dimensional Minkowski spacetime ${\Bbb E}^{3,1}$
by solving the non-linear equation (\ref{eq:Hoppe})for the function $u$
whose level sets are maximal  by separation of variables 
in terms of a certain Weirstrass elliptic function $\frak{p}$.
This construction is very similar to the standard construction of triply
periodic minimal surfaces in ${\Bbb E}^3$ using a different
type of Weirstrass elliptic function \cite{Nitsche}.
He also gives
a quadruply periodic timelike minimal hypersuface. 
Without loss of generality
one may take the invariants $g_2$ and $g_3$ of the Weirstrass function
to equal to $4$ and $0$ respectively. Thus
\ben
(\frak{p}^\prime)^2= 4(\frak{p}^3-\frak{p}).
\een 
The period  is $2 \omega= { 1\over 2 \sqrt{ 2 \pi }} (\Gamma( {1\over 4}))^2$ and the minimum value of $\frak{p}$ is $1$ and  near the origin it has
a double pole with residue $1$. 

The spacelike maximal hypersurface takes the form
\ben
\frak{p}(x) \frak{p}(y)\frak{p}(z)=\frak{p}(t)
\een
and the timelike membrane solution the form
\ben
\frak{p}(x) \frak{p}(y)\frak{p}(t)=\frak{p}(z).
\een
Thus the pure Born-Infeld solution is given implicitly
by
\ben
\frak{p}(x) \frak{p}(y)\frak{p}(z)=\frak{p}(\phi)
\een

Physically the time-independent
solution correpond to an infinite lattice or crystal of Born-Infeld
particles in equlibrium. This seems uncannily appropriate
in view of Born's pioneering work on the the
physics of crystal lattices.

Consider  the solution
inside the cube ${\cal C}\equiv \{(x,y,z) \in 
 [0,2\omega] \times [0,2\omega] \times [0,2\omega] \}$. 
The electrostatic potential $\phi$ takes the
values  $0 \equiv 2\omega$ on the faces of the cube
and the value $\omega$ at the centre $(x,y,z)=(\omega,\omega,\omega)$.
Near the centre one may expand the Weirstrass
function about its minimum value 
to get the approximate form
\ben
(x-\omega)^2 + (y-\omega) + (z-\omega)^2\approx(t-\omega)^2.
\een
Thus the  charges are located at the centres of the
periodic images of the cube $\cal C$.
The maximal hypersurface coincides with the light cone at these
central points. Near the origin we have the approximate
form:
\ben
\phi=xyz.
\een
The potential changes sign as one crosses a face of cube ${\cal C}$
and thus the crystal is of NaCl type, every  charge being surrounded by six
 charges of the opppposite sign,
i.e. the charge at the point $((2n_x+1)\omega,(2n_y+1)\omega,(2n_z+1)\omega),
(n_x,n_y,n_z) \in {\Bbb Z} \times {\Bbb Z} \times{\Bbb Z} $
has the sign
\ben
(-1)^{(n_x+n_y + n_z)}.
\een
Note that $\cal C$ is not what crystallographers call the unit cell.

There are obviously many interesting questions one might ask
about this crystal. For example what are its binding energy
and is its compressibility?
One could study it's electric polarizability by
applying an external electric field. From our previous
work it follows that the relevant solution is obtained
by simply boosting the spacetime solution in the direction
of the applied field. We shall calculate  the binding energy in 
the next section using 
a version of the virial theorem.

The time-dependent solutions  correspond,
by making an electromagnetic duality transformation 
a solution of the 
induction of electric  fields by magnetic field which is periodic in time as well as in two spatial directions. Both are, in the context of Born-Infeld
theory, hitherto unknown and  nicely illustrate the utility
of relating the non-linear Born-Infeld equation to maximal surfaces.

For the Weirstrass functions considered here $1 \over \frak{p}(\tau)$
with $\tau=it$ satisfies the same equation as $\frak{p}(t)$ thus formally
a quadruply periodic minimal solution in ${\Bbb E}^4$
would be  given by
\ben
\frak{p}(x) \frak{p}(y)\frak{p}(\tau)\frak{p}(z)=1.
\een
However the functions considered here have least value $1$ which they
achieve at $0$ and so the euclidean solutions merely consists
of the intersecting hyperplanes  $xyz\tau=0$ and their periodic
recurrences. Thus we get the standard situation of a flat
$4$-brane wrapped
over a torus. This flat
cannot be given a non-trivial electrostatic field 
using the boosting transformation. However
Hoppe's solution can be boosted to give a solution
with non-trivial scalars. One simply replaces $\phi$
with $\phi-vy \over \sqrt{ 1-v^2}$.
The limiting BPS solutions should then
correspond to 
triply periodic harmonic functions.

Hoppe's remarkable Born-Infeld crystal has a 
number of fascinating properties  and provides one of the few many
body solutions. Before discussing them we turn to a general discussion
of multi-solutions.

\section{BIon-Statics }

It is well known that one can construct explicit
multi-black hole solutions
held apart by struts, the struts being the sites
of conical singularities representing ditsributional stresses.
One should be able to construct analogous
multi-BIon solutions. Few  are known explicitly but  
general existence theorems for the Dirichlet problem \cite{Bartnik}show that a solution
exists for each choice of $k$  potentials $\Phi^a$
, $a=1,\dots ,k$ at $k$ prescribed positions  ${\bf x}_a$
of the singularities. One would anticipate physically
that a solution of the dual Neumann problem should also exist if 
one prescribed the charges $q_a$ of the fixed singularities or 
indeed if one
fixed any $k$ dimensional combination of charges and potentials. 
Moreover Pryces's  uniqueness theorem \cite{Pryce2} implies that  
any two solutions 
with charges and potentials $(q_{a 1}, \Phi^a _1)$ and
$(q_{a 2}, \Phi^a _2)$ for which
\ben
\sum (q_{a1}-q_{a2})(\Phi^a _1-\Phi ^a_2)=0
\een 
must in fact be identical. We shall discuss in a later section the forces
necessary to hold the positions of the charges fixed.

This and similar situations in involving 
pinned soliton equlibria 
may be described in terms of a 
canonical formalism. For example the discussion which follows
holds in a wide range of non-linear elctrodynamic theories
with a general Lagrangian.
 
The total  mass is given by the 
integral over $ {\Bbb R}^3 -\{ {\bf x}_a \} $ 
\ben
M= { 1\over 4\pi} \int \bigl ( {\bf E}\cdot{\bf D}-L \bigr) d^3x.
\een
Now let us consider varying the solution
by changing the charges and changing the potentials 
and for the moment keeping the positions fixed.

Using the definition of $\bf D$ we get 
\ben
dM= { 1 \over 4 \pi} \int _{{\Bbb R}^3 - \{ {\bf x}_a \} } {\bf E}\cdot d 
{\bf D} d^3x
\een
\ben
=- { 1\over 4 \pi } \int 
 \nabla \phi \cdot d{\bf D} d^3x.
\een

Now one may use the divergence theorem and the fact that $d {\bf D} $ 
is divergence free to get
\ben
dM =  \sum \Phi ^a d q_a.
\een

In a later subsection we shall obtain an integral relation
for  $M$.
This  resembles the situation of a system of fixed capacitors
in standard linear electrostatics. This resemblance in fact extends
to the extent that a "reciprocity principle" holds. Because the existence
of a reciprocity principle is intimately related to a canonical
or duality invariant formalism almost identical to that in classical 
thermodynamics
 it seems worth while to explore this a little further, not least because
of the possible light it may throw on the subject of black hole 
thermodynamics though we should emphasise from the outset that we are not
at this stage
attempting to ascribe any intrinsic entropy to BIons.

Just as in the conventional case of capacitors one may regard
the potentials as functions of the charges, i.e.
\ben
\Phi ^a = \Phi ^a( q_b)
\een
 or the 
charges as functions of the potentials or indeed one expects to be able to 
take
any $k$ combinations of charges and potentials
as determining the other $k$ variables. In other words the solution set 
is some $k$-dimensional submanifold $\cal L$ of the flat state space
${\cal P} \equiv {\Bbb R}^{2k}$
 with coordinates $(q_a, \Phi^a)$. Note that because our 
problem is non-linear and the principle of superposition
does not hold, the solutions set will not be 
a $k$-dimensional hyperplane as it is in the 
standard case of linear electrostatics.

From the formula for the mass it is clear that
we may think of the state space as the cotangent space ${\cal P} =V \times V^\star$
where $V$ is the vector space of the extensive  charge variables
with coordinates $q_a$ and $V^\star$  is its dual space with 
intensive potential variables $\Phi^a$. The state space $\cal P$ comes
equipped with the symplectic 2-form
\ben
\omega = \sum dq_a \wedge d\Phi^a.
\een

Geometrically the reciprocity relation,
which we will establish shortly, amounts to the assertion that
the solution set $\cal L$ is a Lagrangian submanifold of the
symplectic state space $\cal P$. It is equivalent to the Maxwell
relation
\ben
{\partial \Phi^a \over \partial q_b} =  {\partial \Phi^b \over \partial
 q _a}.
\een
To prove the reciprocity property we subject a solution 
to two independent
variations $\delta _1$ and $\delta_2$ and  integrate over all of 
${\Bbb R}^3$, using the
divergence theorem,
the quantity 
\ben
h= \delta _1 {\bf E} \cdot  \delta _2 
{\bf D}- \delta _2 {\bf E} \cdot  \delta _1 {\bf D}   \een
to obtain the formula
\ben
{ 1\over 4 \pi} \int _{{\Bbb R}^3} h d^3 x= 
\sum \delta_1 q_a \delta _2 \Phi^a- \sum \delta_2 q_a \delta _1 \Phi^a.
\een

However
\ben
h= { \partial ^2 L \over \partial E_i \partial E_j }
 \Bigl ( \delta_1 E _i \delta_2 E_j -\delta_2 E _i \delta_1 E_j \Bigr )=0.
\een
This establishes the reciprocity.

\subsection {Forces and the stress tensor}

We have discussed above static solutions with one or more 
charged particles, possibly in a background  uniform electric
field. The question naturally arises why don't the particles 
accelerate under the influence of the mutual forces? The reason is that
they are pinned to their fixed position ${\bf x}_a$ by external forces.
The existence of this external force shows up in the 
expansion of the solution near the origin 
Comparing the expression for the field
in the presence of an electric field
with  those without an external electric
field we note that the expansions agree to lowest   
order, i.e. the term $-r$ is the same,  but the next term of order 
$r^5$ differs. For the static  single BIon solution without 
an applied field
the order $r^5$ term is isotropic while for the static
BIon solution with an applied elctric field the order $r^5$
term is anisotropic.

The interpretation just given is easily confirmed using
the  conserved and symmetric stress tensor $T_{ij}$.  This is given by
\ben
T_{ij}= \delta _{ij} (L +{\bf B} \cdot {\bf H} ) -E_i  D_j-H_i B _j  = T_{ji}.
\een
By virtue of the static field equation the stress tensor is conserved
\ben
\partial _i T_{ij}=0.
\een
Because we are working in flat Euclidean space 
${\Bbb E}^3$ the total force $F_i$    
acting on a 2-surface $S$ with surface element $d\sigma_i$
is well defined and given by
\ben
F_i= \int _S T_{ij} d\sigma _j.
\een
If one chose a local cartesian frame whose third
leg is aligned with the direction of the gradient of 
the potential function $\phi$ one finds that the stress tensor is diagonal
and has components
\ben
T_{11}=  L
\een
\ben
T_{22}=  L
\een
and
\ben
T_{33}= -2 L ^{\prime} {\bf E} ^2 + L.
\een

Thus  the field
exerts a {\sl pressure} $P= L$ orthogonal to the field  lines,
i.e the direction of the gradient of $\phi$.  
then the field exerts a {\sl tension} 
$- 2 L ^{\prime} {\bf E} ^2 + L$ along the direction of the
field lines\footnote{As long as the Lagrangian function
$L(x)$  satsifies $2xL^\prime(x) -L >0$ this will continue 
to hold in more general non-linear electrostatics}. Of course 
in the case of weak fields
the magnitude of the pressure and the tension are 
equal.

Note that the tension is numerically equal to the Hamiltonian 
function $H({\bf D})$. Thus positivity follows from
the concavity of the Lagrangian function $L({\bf E})$ 
and the properties of the Legendre tansformation.

If $S=\partial \Omega$  is the boundary of a compact domain
 $\Omega \subset {\Bbb E}^n$ one interprets $F_i$ as the 
total force exerted on the material inside the domain $\Omega$
by forces external to $\Omega$. If the fields are everywhere
non-singular inside $\Omega$ then the divergence theorem
implies that the total force on $S=\partial \Omega$ must vanish.  
Conversely if the total force $F_i^{S}$
on a closed surface $S$ is non-zero then it 
must
contain one or more singularities. Since the value of the force
depends only on the homology class of $S$ one may evaluate
the force $F_i^a$ on the $a$'th singularity, assumed pointlike
and finite in number,
inside $\Omega$ by considering a sphere of small radius surounding it.
One then has the identity
\ben
{\bf F} ^{S} = {\bf \sum F}^a.
\een

If $S$ is taken to be a large sphere at infinity and
\ben
\phi \approx -{\bf E}.{\bf x} + {q^{\rm total}  \over r} +
\een
one may evaulate the force and find that 
\ben
{\bf F} ^S={\bf F} ^{\rm total} = q^ {\rm total} {\bf E},
\een
where, of course,
\ben
q^{\rm total} = \sum q^a.
\een

One may  also evaluate the force on 
the singularity by considering a small sphere
about the origin. One may  check explicitly that they agree.
In fact
\ben
F_i={ev \over \sqrt {1-v^2}} \delta _{iz}
\een
but,  as we stated earlier, the 
charge is $q= {e \over \sqrt {1-v^2}}$ and the
electric field is $E=v$ whence
\ben
F_i=qE  \delta _{iz}
\een
as expected.

More generally of the field $\phi$ has a singularity of the form
\ben
\phi = {\rm const} - r + { e r^5 \over 10} + O(r^9) 
\een
,where in general $g$ will be angle dependent, then 
using polar coordinates $(r, \theta, \phi)$ one finds that
the coponents of the force are given by 
\ben
F_1= -{ 1\over 4 \pi} \int \int g^{-{ 1\over 2}} \sin \theta  \cos \phi \sin \theta 
d \theta d \phi 
\een
\ben
F_2= -{ 1\over 4 \pi}\int \int g^{-{ 1\over 2}} \sin \theta  \sin \phi \sin 
\theta d \theta 
d \phi 
\een
\ben
F_1= -{ 1\over 4 \pi} \int \int g^{-{ 1\over 2}} \cos \theta  \sin \theta 
d \theta s \phi .
\een 

If a singularity has a vanishing right hand side it
 experiences no external  force. We  call such singularities
 \lq free\rq. Our usual experience with linear electrodymanics
 encourages the  expectation that in non-linear electrodynamics,
at least if $2xL^\prime(x) -L >0$, then like charges should 
always repel
and unlike charges should always attract. Thus one does not
expect to be able to construct in such theories a solution 
representing $k$ free charges all of the same sign, or 
a solution with two free charges of the opposite sign.
We have seen however that general theorems 
guarantee the existence of solutions
with charges fixed at arbitrary positions
and with no external forces.
It would be interesting therefore to calculate the forces between
the charges for these solutions as a function of separation.
It is not difficult to se that at large separations the forces are 
the standard
inverse square ones but 
unfortunately, exept in the two-dimensional case studied by 
Pryce \cite{Pryce1},
no explicit solutions are as yet availble
to allow one to calculate the forces at close separation.

One thing that is easily done is to rule out
the existence of certain very symmetrical solutions with free charges.
Consider for example two equal and opposite charges at a 
fixed non-vanishing separation. Since there is no external force
coming from infinity, the force between them
could be calculated  by taking the surface $S$ to be the 
plane perpendicular to the line joining the charges
and passing through its  mid point. But by symmetry
the field lines are every where orthogonal to this plane
which is thus a non-compact   level set
of the function $\phi$. Therefore
the total force acting is a tension tending
to attract the charges  in the the direction
of the line and given by the non vanishing integral
\ben
\int _{{\Bbb R}^2} d^2x 2 L ^{\prime} {\bf E} ^2-  L . 
\een
 
If on the other hand the charges have the same sign then
symmetry dictates that the field lines  lie in the plane.
In fact the plane is a degenerate or limiting case of a flux-tube.
A flux tube is by definition a  surface 
of topology of a cylinder or a cone  containing field lines
and thus enclosing a fixed amount of flux of the electric 
induction vector $\bf D$. For a single isolated charge the flux tubes
are right circular cones whose vertices located at 
position of the charge. In the theories we are considering,
flux tubes always experience a pressure
along their normal just as they do in the linear theory.

In any event the  total  force  acting on the plane
is a pressure tending
to repel the charges  in the the direction
of the line and given by the non vanishing integral
\ben
-\int _{{\Bbb R}^2} d^2x  L . 
\een
The relative magnitudes of these two forces
for the same separations and absolute magnitudes of the charges
is not clear but one might expect because the forces should depend on 
the strength of the electric fields $\bf E$ that, at least in 
The Born-Infeld case,
 the repulsion, which tends
to give rise to stronger electric induction fields $\bf D$ 
and hence weaker electric fields $\bf E$
compared with the linear case  is greater 
than the attraction whose effect is in the oppposite sense. However
this intuitive argument is not very conclusive.

The arguments just given, ruling out free charges of equal
absolute magnitudes, could conceivably be extended
to the case when the absolute magnitudes of the charges are no longer equal.
One needs to be able to control the shape of the iso-potentials
or the directions of the field lines or the existence or
and shape of flux tubes.
For example one might be able to replace the separating plane 
in the case of opposite charges by a non-compact level set
of the function $\phi$ separating the charges. This level set
is acted upon by an everywhere by a tension in the direction of its normal.
Thus if the normal always has a positive projection in  
some direction we would be done.
Similar remarks might  apply to the case of charges
of the same sign if if one could control the properties
of the limiting flux tube. Alternatively one might be able to  make use
of suitably chosen planes if one  could control
the direction of the field lines on it.

\subsection{The Virial theorem and the Madelung
constant for 
Bionic cryatals}

If we include a change in the positions of the points
the formula for the  variation in the mass becomes:
\ben
dM= \sum {\Phi}^a dq_a + {\bf F}^a \cdot d {\bf x}_a .\label{eq:First}
\een
We could obtain this by
a detailed variational calculation
using the fact that the canonical
stress tensor is related to a variation
of the energry density with respect to postion,
but in the presnt case
it is easier to derive it by elementary 
dimensional analysis. We start by deriving
the virial theorem, which is  essential an integrated version of this
identity by   integrating over ${\Bbb R}^3$ the identity
\ben
\bigl (T_{ij}x_k \bigr ) ,_i= T_{jk}
\een
Contraction over $ij$ and use of boundary conditions,
 the formula for the trace
\ben
T_{ii}={\bf E} \cdot {\bf D} -3L
\een
and the formula for the mass gives the Smarr-type relation,
\ben
M= {1 \over 3} \sum {\bf F}^a \cdot{\bf x}_a + { 2 \over 3} \sum q_a {\Phi ^a}
\label{eq:Smarr}.
\een
This formula, for a single spherically symmetric charge,
with no applied force was known to Born\cite{B}. It is interesting to 
compare
it with the Smarr formulae one obtains for black holes.
This is also a consequence of a virial type theorem.

Note that if we were working in $p$ spatial dimensions
the variational formula ({\ref{eq:First}) remains true but
the virial theorem (\ref{eq:Smarr}) would be:
\ben
M= {1 \over p} \sum {\bf F}^a\cdot{\bf x}_a + { p-1 \over p } \sum q_a {\Phi ^a}
\label{eq:Smarrp}.
\een

Now if $\phi({\bf x};{\bf x}_a)$ is a solution of the static Born-Infeld equations with singularities at the
points $\{{\bf x}_a\}$ having total mass $M$, charges $q_a$,
potentials $\Phi^a$ and forces $F^a$
then $\lambda ^{-1}(\lambda {\bf x}; \lambda {\bf x}_a)$
will also be a solution with singularities at $\lambda {\bf x}_a$
total mass $\lambda ^{-p} M$, charges $ \lambda ^{-(p-1)} q_a$,
potentials $ \lambda ^{-1}  \Phi^a$ and forces $ \lambda ^{-(p-1)}  {\bf F}^a$.
The passage between (\ref{eq:First}) and (\ref{eq:Smarrp})
is a trivial application of Euler's theorem.
Note that the scaling relation (\ref{eq:scaling}) used
by \cite{DLS} in their discussion of fission 
is a special case of this general scaling relation.

For a crystal the integral over  ${\Bbb R}^3$ will not
converge but one can integrate over the cube 
${\cal C}=[0,2\omega] \times [0,2\omega]\times [0,2\omega]$.
In the case of the Hoppe solution the potential vanishes 
on the walls  of the cube and takes the value $\omega$
at the single charge located at its centre. It is clear 
by symmetry that that the force on the charge vanishes
and therefore
the energy per unit volume is 
\ben
{ E\over V}= { 1 \over 8 \omega^3}  { 2\over 3} \omega e,
\een
where $e$ is the value of the electric charge.
Note that if we scale the solution by a fcator of $\lambda$ 
the size of the unit cell, the electrostatic potential
and the energy will all scale but the ratio $ { E\over V}$
is unchanged. In other words if we restore
units the energy density is a multiple of $(2 \pi \alpha ^\prime)^{-2}$.

The charge may be found by expanding the elliptic functions
about the point $(\omega, \omega, \omega)$.
Now
\ben
t-\omega= { 1\over 2} \int ^{\frak p(t)}_0 { dh \over {\sqrt{h(h^2-1)}}},
\een
and the half-period is given by
\ben
\omega
= { 1\over 2} \int ^\infty_0 { dh \over {\sqrt{h(h^2-1)}}}.
\een
As a check one sees that  near $t=2\omega$
\ben
{\frak p} \approx { 1 \over (t-2\omega )^2}
\een
as expected.

Expanding the integral gives:
\bea
t&=&\pm  \sqrt{{ ({\frak p}(\omega+t)-1) \over 2}} \Bigl( 1- { {\frak p}(\omega+t)-1\over 4} \nonumber \\
 &+&{ 19 \over 160} { ({\frak p}(\omega+t)-1)}^2  - 
{ 9 \over 128} { ({\frak p}(\omega+t)-1)}^3 \Bigr ) \nonumber \\
 &+& O( { ({\frak p}(\omega+t)-1) }^{ 7\over 2} ), \nonumber \\ 
\eea
and reversion of this series gives:
\ben
{\frak p}(\omega +t)= 1+2t^2 + 2 t^4 + { 8\over 5} t^6 + + { 6\over 5} t^8
+{64 \over 75} t^10 + O(t^{10}).
\een
Near $(\omega, \omega, \omega)$
we therefore have:
\begin{eqnarray}
&&{\frak p}( \omega +x) {\frak p }( \omega +y) {\frak p}( \omega +z)-1 =  \nonumber \\
& & 2r^2 + 2 r^4 + 8x^2  y^2 z^2  + { 8 \over 5} ( x^6 +y^6 + z^6 ) \nonumber \\
& & + 4( x^4 y^2 + x^4 z^2 + y^4 z^2 + y^4 x^2 + z^4 x^2 + z^4 y^2 ) + O(r^8).
\nonumber \\
\end{eqnarray}
Thus, finally, 
\ben
t=\omega - r +{ (x^2+y^2)(y^2+z^2)(z^2+x^2) \over 5 r} + O(r^9),
\een
where we have resolved the ambiguity in the square root by considering
taking the charge to be positive. Note that there
is no $O(r^7)$ term as expected on general grounds and that
the symmetry of the $O(r^5)$ term implies that the force
does indeed vanish.

The charge is given by  the spherical average of 
$1 \over \sqrt {2(x^2+y^2)(y^2+z^2)(z^2+x^2)}$.
One could compare the self-energy with the Madelung constant
for an NaCl structure. Formally the potential energy of the
positive charge in the field of the other charges is
\bea
&&-{e^2 \over 2\omega} 
\Big( 6- {12 \over \sqrt{2}} + { 8 \over \sqrt{3} } 
\dots \Bigl )\nonumber \\
&&\approx - 1.748 {e^2 \over 2\omega}. \nonumber \\
\eea
Having demonstrated how to calculate 
crystal energies in Born-Infeld theory we will postpone to 
another occasion the detailed numerical comparison.
\footnote{ Strictly speaking the series in brackets, $
\sum { (-1)^{n_x+n_y+n_z} \over \sqrt 
{ n_x^2 + n_y^2+ n_z^2 }}~~
(n_x,n_y,n_z) \in {\Bbb Z}\times {\Bbb Z}\times {\Bbb Z} \setminus (0,0,0)
$,
is not convergent. 
One way \cite{madelung} of regularizing is to
define 
$
d_3(2s)=\sum { (-1)^{n_x+n_y+n_z} \over (  n_x^2 + n_y^2+ n_z^2 )^s}~~(n_x,n_y,n_z)\in {\Bbb Z}\times {\Bbb Z}\times {\Bbb Z} \setminus (0,0,0),
$
for ${\rm Re} s>{3 \over 2}$ and analytically continue.
One gets the same answer.}

\section{Catenoidal solutions}

As mentioned earlier there are some non-trivial
static solutions with vanishing gauge fields.
The equations of motion on ${\Bbb E}^p$ are
\ben
\nabla \cdot { \nabla y \over \sqrt{ 1+ |\nabla y|^2 } }=0.
\een
There is obviously an  $SO(p)$-invariant solution such that
\ben
{ \nabla y \over \sqrt{ 1+ |\nabla y|^2 }}=  {{ \bf x} \over r^p}.
\een
but unlike
the electrostatic solution it is not defined everywhere outside 
the  origin. Since
\ben
\partial _r y= \sqrt { (r^{2p-2}  -1)} 
\een
Thus   $y$ has a square root singularity at 
\ben
r=1.
\een
Therefore one should  regard $y$ as double valued and
branched over  the $p-1$-sphere at this critical radius.
Geometrically this means smoothly joining
the original incomplete  minimal surface to another identical one
to obtain a smooth catenoidal solution. The critical
$p-1$ sphere then becomes a minimal hypersurface lying
in the catenoid. If  $p>2$ $y$ tends to a finite limit
as the radius tends to infinity and so such a catenoid looks like 
two parallel asymptotically planar surfaces a finite distance apart
joined by a throat. 
The case of familar experience, $p=2$, is exceptional in that $y$
does not tend to a finite limit as the radius tends to infinity.
Thus if $p=2$ the two ends are infinitle far apart and never really become
planar. We shall give a  global emebedding 
of the catenoid in the next subsection. 
These catenoids are strikingly similar to the
Einstein-Rosen bridges, i.e. the
surfaces of constant time, that one encounters in
classical super-gravity solutions representing black holes or
black $p$-branes. We shall see shortly that this analogy
goes even deeper.

It is natural to  ask whether static
multi-catenoidal solutions exist.
One might doubt it  because the scalar field $y$ should be responsible
for a long range attraction bewtween catenoids. The
easiest way to check this is again to calculate the stress
tensor. For a pure scalar solution one has
\ben
T_{ij} = \delta _{ij}L + { 1\over 1-L} \partial_i y \partial_j y. 
\een
Thus if one had a symmetrical configuration which is 
symmetrical about the plane $x^3=0$, then one would have
$\partial _3y=0$ there. It follows that $T_{33}$
would be negative. This means, as expected. that there is a net
attractive force and thus equilibrium between two
symmetrical catenoids should not possible. However
this argument cannot be extended in a straightforward 
fashion to more complicated configurations 
with a finite number of throats
and in the
light of the discoveries 
of recent years of many (unstable)
minimal surfaces in ${\Bbb E}^3$ \cite{Hoffman,Nitsche} caution is advisable.

One can be less cautious about periodic arrays.
The solutions exhibited by Hoppe et al. definitely
show that such solutions exist.

\subsection{Cosmological and Wormhole solutions}

Since the catenoid and the BIon solution
play such an important role
in the theory we shall conclude this section with a 
few additional remarks about
their geometry and that of related solutions.
As was remarked earlier, the flat $p+1$-plane corresponds to the trivial
Minkowski vacuum state from the point of view
of physics on the brane. There is in fact a braney
cosmological solution  given (in the case of p=3)
 in \cite{cosmos}
\footnote{Note that, in contrast to the considerations of \cite{cosmos},
 since there is no gravity on the brane
there is no question of applying Einstein's equations
on the brane  in the present context}.
The metric induced on the world volume
corresponds to an F-L-R-W metric 
\ben
ds^2= -dt^2 + a^2(t) d \Omega^2 _{p,k}
\een
where $\Omega^2 _{p,k} $ is the metric on a unit
 $p$-sphere if $k=1$, the euclidean metric if $p=0$ and the metric on hyperbolic space if $k=-1$. 
If $k=1$ the embedding is given by
\ben
Z^i= a(t) \cos \chi n^i 
\een
\ben
Z^{p+1}=a(t) \sin \chi  
\een
\ben
Z^0 = \int \sqrt{1- \dot{a} ^2 }dt.
\een
where $n^i$ is a unit vector in ${\Bbb R}^{p}$.
The equations of motion reduce to
a Friedmann type equation
\ben
{ \dot{a}^2 + 1}= { 1 \over a^{2p} }.  
\een  
where a convenient choice of the arbitrary length scale
has been made.

The big-bang and big crunch singularities
 where $a(t)\rightarrow 0$
corresponds to the Ecker-type of singularities\cite{E} 
where $p$-brane world sheet becomes tangent to the light cone
through the origin of ${\Bbb E}^{p+1,1}$.

The catenoid solution is obtained by analytic continuation
setting $t=i\tau$ and $Z^0=iy$ where
$\tau$ and $y$ are real. The static gauge 
consists of using the coordinates $Z^i, Z^{p+1}$ as
coordinates and thus $a$ is to be identified
with the radial coordinate used earlier.
The analytically
continued Friedman equation
then becomes
\ben
{ -{a^\prime}^2 + 1}= {1 \over a^{2p}}
\een
where $^\prime$ denotes differentiation with respect to $\tau$.
If one solves this one finds that $a(\tau)$ is defined for all $\tau$.
It is symmetrical about its unique minimum value
and tends to $|\tau|$ as $|\tau| \rightarrow \infty$.
Thus solving for $y$ as a function of $r$
gives two valuess if $r$ is larger than a critical value
and none if it is smaller. This is precisely the behaviour we found earlier
and it is a simple matter to check that the explicit formula for $y(r)$
obtained from the Friedmann equation is the same as that obtained earlier.

One may also obtain the Bion solution
as a spacelike maximal surface in this way. We set
\ben
Z^0 = \int \sqrt{-1+ \dot{a} ^2 }dt.
\een
and the Friedmann equation becomes
\ben
{ \dot{a}^2 -1 } = { 1 \over a^{2p}}.
\een
As before one identifies  $Z^0$ with 
the electrostatic potential $\phi$.

\subsection{Energetics of catenoids}

One may define the total energy $M$ of the static
catenoid
solution with respect to one of its sheets by
\ben
M= \int T_{00} d^p x=\int ( \sqrt{ 1 + |\nabla y|^2 }-1) d^px,
\een
where the integral is taken over the region outside the throat.
The equation of motion implies
that one may associate with the solution a  \lq scalar charge\lq 
\ben
\Sigma =  -\int {\nabla y \over \sqrt{ 1+ |\nabla y|^2} } \cdot d \sigma
\een
where the integral may be taken over any $S^{p-1}$ surrounding the throat.
Multiplying the equation of motion by $y$ and integrating over the 
region outside the throat gives
\ben
\int  {|\nabla y|^ 2 \over \sqrt{ 1+ |\nabla y|^2} } d^px =Y \Sigma,
\een
where $Y$ is the difference between the value 
of $y$ at the throat and its value at infinity. 
Thus the distance between the two 
asymptotically flat regions  is $2Y$. 
We have chosen conventions so that both $\Sigma $ and $y$ may be taken
to be positive.

Now the virial theorem is more subtle because the boundary term
at the throat must be taken into account.
One has
\ben
T_{ii}= {|\nabla y |^2  \over \sqrt{ 1+ |\nabla y|^2} }+p( 1-\sqrt{ 1+ |\nabla y|^2}).
\een
One  has
\ben
\int rT_{rr} d \sigma_r=
Y \Sigma Y-pM.
\een
The boundary term works out to be $-pV$\footnote{
the reader is cautioned against confusing $p$ as in $p$-brane
with $P$ as in pressure.}
 where $V$ is the excluded volume
and therefore
\ben
pM= Y\Sigma +pV.
\een

As before we may consider a family of solutions of the form
$\lambda ^{-1}y(\lambda {\bf x})$ for which
the mass and volume scale together  like $\lambda ^{-p}$, 
$Y$ like $\lambda ^{-1}$ and $\Sigma$ like $\lambda^{-(p-1)}$. 
The variational
formula becomes
\ben
dM= dV + { 1\over p} \Sigma dY.
\een

If one  thinks of $2M$ as the total energy of the 
catenoid and we recall that $2Y$ is the distance bewteen the two ends
one might be tempted to
regard  $ { 1\over p} \Sigma$ as the tension in the throat.

\section{Born-Infeld-Electrostatics with extra scalars}

In this section we shall consider solutions when both the scalar fields
and the electric field are non-zero. 
The Lagrangian is
\ben
-\sqrt{  -{\rm det} (\eta_{\mu \nu} + F_{\mu \nu} + 
\partial _\mu y^n \partial_\nu y^n )}.
\een
If we make the electrostatic ansatz we need to evaluate 
\ben
\left | 
\pmatrix {
{-1} & {-\partial _i \phi }\cr 
{\partial _j \phi} & {\delta _{ij} + \partial _i y^n \partial_j y^n }\cr 
} 
\right |. 
\een
 
On use of the  matrix identity one finds that one needs
\ben
\sqrt {{\rm det} ( \delta _{ij} - \partial _i \phi \partial _j \phi 
+\partial_i y^n \partial_j y^n )}.
\een
Thus one is concerned with a  spacelike $p$ dimensional maximal
hypersurace in  $d+1$ dimensional Minkowski spactime ${\Bbb E}^{d,1}$.
Of course if $\phi=0$ we get minimal $p$-surface in ${\Bbb E}^d$.
The simplest case in which one can evaluate the determinant is when $d=p+2$
and so there is just one scalar $y$. The result is that one must extremize
\ben
\int d^px \sqrt { 1 + |\nabla y|^2 -|\nabla \phi |^2 + (\nabla y\cdot \nabla \phi)^2 - |\nabla \phi|^2 |\nabla y|^2 }.
\een

Note that in accord with our general results one may indeed
consistently set
$y=0$ or $\phi=0$.

\subsection{Charged Catenoids}

We have seen that the Dirac-Born-Infeld equations have 
pointlike solutions with vanishing scalar and with
finite energy which exhibit the expected attractions and repulsions. 
We also have the pure scalar catenoidal solutions with
no electromagnetic field which are attractive. 
None of these solutions can therefore not be expected to be supersymmetric.
However as we shall see
there is a simple ansatz which does give rise to multi-centred
solutions 
depending
on an arbitary harmonic function. 

The static equations of motion require the vanishing of
\ben
\nabla \cdot { -\nabla \phi +\nabla y (\nabla y \cdot \nabla \phi) - \nabla \phi
(\nabla y)^2 \over
\sqrt{ 1-(\nabla \phi)^2 + (\nabla y)^2 + (\nabla y \cdot \nabla \phi)^2
- (\nabla y)^2 ( \nabla \phi)^2 }}=0 
\een
and
\ben
\nabla \cdot  { \nabla y +\nabla \phi (\nabla y \cdot \nabla \phi) - \nabla y
(\nabla \phi)^2 \over
\sqrt{ 1-(\nabla \phi)^2 + (\nabla y)^2 + (\nabla y \cdot \nabla \phi)^2
- (\nabla y)^2 ( \nabla \phi)^2 }}=0. 
\een
Note that the electric induction $\bf D$ gets a contribution from the scalar
field
\ben
{\bf D}= 
{
 -\nabla \phi +\nabla y (\nabla y \cdot \nabla \phi) - \nabla \phi
(\nabla y)^2 
\over
\sqrt{ 1-(\nabla \phi)^2 + (\nabla y)^2 + (\nabla y \cdot \nabla \phi)^2
- (\nabla y)^2 ( \nabla \phi)^2} 
}
\een

The manifest invariance under boosting
in the $\phi-y$ variables gives rise to  a solution transformation
 which is completely analogous to the well known Harrison transformations
in Black Hole theory which allow one to pass from neutral black hole
to a charged black hole by a boost. Specifically if $(\phi_0,0)$ 
is any pure Born-Infeld
electrostatic field with no scalar $y$ then 
$({1 \over \sqrt{ 1 -v^2}} \phi,{1 \over \sqrt{ 1 -v^2}} \phi)$ 
are also solutions of the Dirac-Born-Infeld equations of motion. If $\phi$ has finite energy then so will the new solution. 
Obviously we can apply the same procedure to a solution
$(0,y)$ with scalars but no vectors. $({1 \over \sqrt{ 1 -v^2}}y ,{v \over \sqrt{ 1 -v^2} }y )$ will also be a solution.

Geometrically (regarding $\phi$
as an electrostatic potential rather than a transverse coordinate)
one has a non-singular source-free deformation of the catenoid solution
in which the distance between the two asymptotic $p$-planes in increased
by  a factor $ {1 \over \sqrt{ 1 -v^2}}$. More significantly
the catenoid is now electrically charged. This is only compatible
with the absence of singularities or sources because the
catenoid solution is topologically non-trivial. The electric field
lines can thread through the minimal $p-1$ cycle separating the two
asymptotic regions. The throat or tunnel clearly becomes
more and more narrow and string-like as one approaches extremality.
This is strikingly reminiscent
of the behaviour of sub-extreme black holes and black branes.
The solutions obtained by boosting the electrical BIon
solution is of course analogous to the behaviour
of super-extreme black holes and black branes.

\section {Electric BPS solutions}

The natural question is what happens
in the limit as  $v\rightarrow 1$? The answer is 
that the non-linear equations linearize and one finds that
one can solve the equations my making the ansatz that
\ben
\phi = \pm y= H
\een
and then discover  that $H$ may be an arbitrary harmonic function on 
${\Bbb E}^p$.
Choosing $H$ to be a sum of simple poles
gives Coulomb like solutions  representing
point particles with infinite total energy\footnote{The spherically symmetric solution was found first in \cite{DLS}.}.
Note that if $\phi$  is interpreted as an extra 
timelike coordinate then the  resulting maximal
surface in ${\Bbb E}^{p,1}$ is a null hypersurface
with lightlike normal ${ \partial \over \partial \phi} \pm { \partial \over \partial y}$.  

If one regards $\phi$
as an electrostatic potential rather than a transverse coordinate
the distance beween the two asymptotic $p$-planes of the deformed catenoid 
tends to infinity, since $y$ tends to infinity at the origin.

The  supersymmetry transformations of 10-dimensional
Yang Mills theory are of the form \cite{DLS}
\ben
\delta \lambda = ( \gamma ^{\mu \nu } F_{\mu \nu} + \gamma^{\mu m} \partial_\mu  y^m) \epsilon.
\een
It is easy to see that one picks $\epsilon$ 
such that $(\gamma ^0 \pm \gamma ^y) \epsilon=0$
then $\delta \lambda=0$ if the lightlike condition $\phi=\pm y$
is satisfied. Thus our Coulomb solutions are indeed BPS
for all $p$.
A single BPS solutions may be thought of as the limit of 
two $p$- branes  as the separation tends to infinity.

\subsection{ Inclusion of a magnetic field}
An evaluation of the relevant determinant
yields the Lagrangian
\bea
&&L= 1-\nonumber \\
&& \sqrt{ 1 + |\nabla y|^2 -|\nabla \phi |^2 + (\nabla y\cdot \nabla \phi)^2 - |\nabla \phi|^2 |\nabla y|^2  
+{\bf B}^2  -({\bf B} \cdot \nabla \phi)^2 + ({\bf B} \cdot \nabla y))^2
}.\nonumber \\
\eea
Note that despite its greater complexity
the Lagrangian remains invariant under
the global action of boosts, i.e. of $SO(1,1)$ 
on the fields $\phi$ and $y$
provided ${\bf B}$ is not transformed. Moreover we
know that the equations are invaraint under the $SO(2)$ 
duality symmetry. Thus one might anticipate that
the combined equations have an $SO(2,1)$ symmetry.
As we shall see this does indeed turn out to be the case.

As always,
the Lagrangian has to be varied subject to the constraint that
the magnetic induction is divergence free
\ben
\nabla \cdot {\bf B}=0,
\een
and leads to the equation of motion that the magnetic field
\ben
{\bf H} =-{ \partial L \over \partial {\bf B}}
\een
is curl-free:
\ben
\nabla \times {\bf H}=0.
\een
As in the case with no scalar, it is convenient
to perform Legendre transformation.
After some algebra one finds the 
expression
\bea
&&{\tilde H}= 1- \Bigl ( 1- (1-{\bf E}^2)( 1-{\bf H}^2) ( 1 +(\nabla y)^2)
+ ( 1-{\bf E}^2)(\nabla y \cdot {\bf H})^2\nonumber \\
&& -( 1+ (\nabla y)^2 ) ( {\bf E}\cdot {\bf H})^2 + ( 1-{\bf H}^2) ( {\bf E}\cdot \nabla y )^2 + 2({\bf E} \cdot \nabla y)( {\bf E}\cdot {\bf H}) ( {\bf H} \cdot \nabla y) \Bigr )^{ 1\over 2} .\nonumber \\ 
\eea
Expressed in terms of the electrostatic and magnetostatic potentials 
and the scalar $\Phi^S= (y, \phi,\chi)$ and three-dimensional
Minkowski metric $\eta ^{RS}$ 
with the  convention that $\eta ^{yy}=1$, it takes the 
manifestly $SO(2,1)$ -invariant form 
\ben
1-\sqrt {{\rm det} \Bigl (  \nabla \Phi^R \cdot  \nabla \Phi ^S  -\eta ^{RS}\Bigr)}. 
\een
Using the general matrix identity
\ben {\rm det} ({\Bbb I}_n + AB ) ={\rm det} ({\Bbb I}_m+ B A )
\een
for an $n \times m$ matrix $A$ and an $m \times n$ matrix $B$
one may show that
\ben
{\rm det}  \bigl ( g_{\mu \nu} +  \partial _\mu y^m  G_{mn} \partial _ \nu y^n \bigr  )
= { \rm det} g_{\mu \nu} ~{\rm det} G_{pq}~ {\rm det} ( G^{mn} + \partial _\mu y ^m g^{\mu \nu} \partial _\nu y^n ).
\een

It follows that $(y, \phi, \chi)$ extremize
the   Dirac action in static gauge for a 
static 3-brane 
in ${\Bbb E}^{5,2}$. In other words the electric and magnetic potentials
may be thought of as extra timelike coordinates.

\subsection{ Magnetic Monopoles}

If $p=3$ one may find another supersymmetric set of BPS solutions
by duality rotations. The simplest case correponds to solutions
of the  abelian Bogomol'nyi solutions for which
\ben
{\bf B}= \pm \nabla y,
\een
or in terms of the magnetic potential
\ben
\chi= \pm y.
\een

The transverse coordinate $y$ now plays the role of a Higgs
field.

These solutions may be thought of as infinitely separatated branes
since the coordinate $y$ tends to infinity at the centre.
It is expected \cite{W} that one the non-abelian  $SU(2)$ theory
results If one passes to opposite limit of two coincident $p$-branes.
The intermediate case is believed to correspond
to a spontanously broken $SU(2)$
Yang-Mills theory  with the Higgs field in the adjoint
representation. It is natural to identify the $3$ component
of the Higgs with the scalar $y$.
The mass of vector bosons correponding to 
$W^\pm _\mu= { 1\over \sqrt 2} \Bigl (A^1_\mu \pm i A^2_\mu \Bigr )$
is proportional to the separation of the two branes. As the separation
tends to infinity we should therefore recover the abelian theory and this is consistent with what we have found. It also strongly
suggests that interesting Bogolmolnyi type
solutions  exist in non-abelian Yang-Mills theory (see \cite{NS}).

Acting with $SO(2,1)$ one can obtain the general  Dyonic
BPS solution which corresponds to the light cone
in $\Phi^R=(\phi,\chi,y)$ space. The mysterious phase in the 
moduli space of non-abelian BPS monopoles is just
the circle of light-like directions.

\subsection{Bogolmol'nyi Considerations}

In the  energy density is
\ben
T_{00}= {\bf E}\cdot {\bf D}-L
\een
This turns out to be
\ben
{1+ (\nabla y)^2 \over \sqrt { 1+|\nabla y|^2 -|\nabla \phi |^2
+ (\nabla y\cdot \nabla \phi) ^2 - (\nabla y)^2(\nabla \phi)^2 }}-1.
\een
As long as $|\nabla \phi|^2 \le |\nabla y|^2$,
the energy density is bounded below by 
\ben
 (\nabla y)^2.
\een
One may calculate the spatial stress tensor by coupling
the Lagrangian to a background metric $\gamma _{ij}$ 
and taking a variational derivative of ${\cal L}=\sqrt \gamma L$
where $\gamma = {\rm det } \gamma _{ij}$. 
\ben
-T_{ij} = { 2 \over \sqrt{ \gamma} }{ \delta {\cal L} \over \delta \gamma ^{ij}}.
\een
and thus
\ben
-T_{ij}=  2 { \delta { L} \over \delta \gamma ^{ij}} -\gamma _{ij} L .
\een
Now $-L-1$ is given by
\ben  
\sqrt{ 1 - \gamma ^{ij}( \partial \phi \partial _j \phi -   \partial y \partial _j y ) + \gamma ^{ij} \gamma ^{pq}
( \partial _i \phi \partial _j y \partial_p \phi \partial _q y -
\partial _i \phi \partial _j\phi  \partial _p y \partial _q y )
 }.
\een
Therefore
\bea
&&T_{ij}=\gamma _{ij}L +{ 1\over 1-L} \bigl ( \partial _i y \partial _j y
- \partial _i \phi \partial _j \phi  + 2( \partial _{(i} \phi \partial _{j)}
y ( \nabla \phi \cdot \nabla \phi) \nonumber \\
&&
- \partial _i \phi  \partial _j \phi   
( \nabla y) ^2 -\partial_i  y \partial _j y (\nabla \phi )^2 \bigr ) .\nonumber \\
\eea

Evidently the stress tensor vanishes pointwise in the BPS limit
$y=\pm \phi$. 
By contrast if one considers the symmetrical situation
envisaged earlier in which a particle and an anti-particle lie
in the 3-direction then on the symmetry plane $\partial _3\phi=0=\partial _3 y$ and hence 
\ben
T_{33}= L.
\een
It follows by a repetition of the analysis given above that
as long as they are super extreme,
the particle and anti-particle attract.
 
\section {Calibrated p-branes}

In this section we shall show that the idea that strings ending on $p-branes$ or throats connecting two branes can be interpreted as topological
defects on the brane applies more generally.
To see this we note that one supply of minmimal submanifolds,
is provided by using \lq calibrations\rq \cite{HB}.
These satisfy a Bogomol'nyi  like-property
that minimize  volume among all 
submanifolds  in the same  homology class. Thus they have
a good chance of being supersymmetric. Particular cases
are special Lagrangian submanifolds and Cayley surfaces
These correspond to supersymmetric cycles \cite{Beck}.

\subsection{Lagrangian Submanifolds Global Monopoles and Vortices}

In this subsection we consider special Lagrangian submanifolds of 
${\Bbb C}^p \equiv {\Bbb R}^p \oplus i{\Bbb R}^p$\cite{HB}.
If $({\bf x},{\bf y})$ are coordinates for ${\Bbb R}^p \oplus i{\Bbb R}^p$
then these take the form in static gauge
\ben
y_i={\partial}_i F(x)
\een
where  the generating function $F(x)$ satisfies
\ben
{\rm Im}~~ {\rm det} (\delta_{ij} +{\sqrt{-1} ~ \partial _i \partial _j F )}
=0.
\een
Harvey and Lawson give some $SO(p-1)$-invariant examples with  topology
${\Bbb R} \times S^{p-1}$.
The solutions have
\ben
{{\bf y} \over |{\bf y}| } = { {\bf x} \over |{\bf x}| },
\een
\ben                                                                         {\rm Im}  ~~(|{\bf x}| +i|{\bf y}|)^p=c,
\een
where $c$ is a constant
and so they resemble global monopoles. If $c$ vanishes
we have a set of $p$-planes meeting at the origin and making  an
angle $\theta$ such that  $\sin (p\theta)=0$.
If $p=3$ 
\footnote{If p=3 then $\sqrt{ {\rm det}(\delta_{ ij}+\partial_i {\bf y}\cdot \partial_j {\bf y})}$  may be re-interpreted
as the energy function of a rather unusual
non-linear super or hyper elastic material with rubber-like properties.}
we have
\ben
3r^2=  y^2+ {c \over y}.
\een
If $c$ is positive it is convenient
to  choose the arbitrary
scale so that $c=2 $. The solution,
are considered as a function of $r$. It is defined only if
$r> 1$. The solution is double valued  for $r>1$. One branch
has 
\ben
{\bf y}  \approx { {\bf x} \over 6 |{\bf x} |^3}
 ~~~~~~~~~{\rm as }~~~~r \rightarrow \infty
\een
and behaves exactly  like a global monopole in a
theory with a global  $SO(3)$ symmetry \cite{Barriola-Vilenkin}.
The other branch is linear and behaves like 
\ben
{\bf y} \approx \sqrt 3 {\bf x}~~~~~~~~~{\rm as }~~~~r\rightarrow \infty.
\een
Geometrically the 3-brane looks like two 
asymptotically planar regions 
in ${\Bbb E}^6$ joined by a throat. The two
asymptotic 3-planes intersect at the origin 
making an angle of ${\pi \over 3}$. 
 
If $p=4$ the 4-brane  also two branches and the asymptotic
4-planes make an angle of
$\pi \over 4$. If one sets $c=4$, one has
\ben
r^3y-ry^3 = { c \over 4}
\een
and hence one branch is monopole like with
\ben
{\bf y} \approx { {\bf x} \over |{\bf x}|^4}
 ~~~~~~~~~{\rm as }~~~~r\rightarrow \infty
\een
while the other branch is linear
\ben
{\bf y} \approx  {\bf x}~~~~~~~~~{\rm as }
~~~~r\rightarrow \infty.
\een
If $p=5$ one has 
\ben
y(5r^4-10r^2 y^2 +y^4)=c.
\een
There are now two types of interesting solution.
One has an  asymptotically monopole-like branch with
\ben
{\bf y} \approx { {\bf x} \over |{\bf x}|^5 }
 ~~~~~~~~~{\rm as }~~~~r \rightarrow \infty
\een
joined to a linear branch with
\ben
{\bf y} \approx \sqrt {5-2\sqrt 5} ~~{\bf x}~~~~~~~~~{\rm as }
~~~~r \rightarrow \infty.
\een
The other type of solution has two linear branches with
\ben
{\bf y} \approx \sqrt { 5 \pm 2\sqrt 5 }~~{\bf x}~~~~~~~~~{\rm as }
~~~~r \rightarrow \infty
\een
joined by a throat.

Using the boost invariance of the Lagrangian
it is possible to give these solutions a Born-Infeld
electrostatic static field. The resulting electric fields are then
of dipole character.

If $p=3$ the generating function $F({\bf x})$ of a special Lagrangian 
submanifold satisfies:
\ben
\nabla ^2 F = {\rm det} ( \partial_i \partial _j F)\label {eq:hess}.
\een
One may ask for solutions for which both sides of (\ref{eq:hess}) vanish.
The vanishing of the right hand side  is the condition that the level sets
of $F$ be developable. Thus we are looking for developable surfaces 
which are also harmonic. Not surprizingly the helicoid:
\ben
F=a z+ b \arctan y/x
\een
provides a solution. It corresponds to a vortex on the brane
\bea
y^1 &= &-a{y \over x^2 + y^2 } ,\nonumber \\
y^2&=& a{x \over x^2 + y^2 },\nonumber \\
y^3&=b&.\nonumber \\
\eea
By a theorem of Catalan \cite{carmo}, the helicoid
is the only ruled minimal surface in ${\Bbb E}^3$ other than the plane.
As we have seen above, by  Hamel's theorem the 
helocoid is the only harmonic minimal surface in ${\Bbb E}^3$
other than the plane. It is natural to conjecture
that the helicoid is the only harmonic developable in ${\Bbb E}^3$ 
other than the plane.

\subsection{ A Cayley example}

Another interesting example  of a similar general
character given by Harvey and Lawson \cite{HB},
resembles certain types of \lq textures \rq
modelled on the Hopf fibration. 
It is  a Cayley 4-brane living in ${\Bbb E}^7\equiv {\rm Im}  {\Bbb O}
\equiv {\Bbb H} \oplus  {\rm Im} {\Bbb H} $. The 4-brane may be exhibited
in static gauge as a graph over ${\Bbb R}^4 \equiv {\Bbb H}$  
by giving the height 3-vector ${\bf y}$ as a function of a  
quarternion coordinate
$q$. One has
\ben
{{\bf y} \over |{\bf y}|} = { \sqrt 5 \over 2 }{ q i {\bar q} \over |q|^2},
\een
with
\ben
y(4 y^2-5r^2)=c.
\een
Recalling that $r=|q|$ we see that if the constant $c$ is positive then
the solution is only defined outside a critical 3-sphere whose value
depends on $c$  and is double valued outside this critical
throat or 3-sphere. One branch is asymptotically flat with
\ben
y \approx {c \over 25 r^4}.
\een
The other branch tends to the cone 
\ben
{\bf y} \approx {\sqrt 5 \over 2} {q i {\bar q} \over |q|}. 
\een
The  cone is not asymptotically flat and this respect
there is some resemblance to certain super gravity
solutions with $G_2$ holonomy which are also not asymptotically flat
\cite{duff}.

\section{Acknowledgements}

The material in this paper owes a great deal to discussions 
and correspondence with many people.
They are far too numerous to enumerate  in detail here. 
However I would particularly like to thank Robert Bartnik 
whose advice about maximal surfaces and their
literature was extremely valuable at an early stage. 
Conversations with John Schwarz, David Lowe and  Malcolm Perry
were extremely useful, allowing me to see more clearly
the relevance of scalars, and discussions with Costas Bachas,
 Michael Green,
George Papadopoulos and Paul Townsend on D-branes
were extremely helpful.
Moti Milgrom's told me about his independent 
and unpublished work on inter-particle forces in classical field theories
which resembles closely that in section (4.1) 

During the fnal stages of writing up
I became aware of the work of 
\cite{callan}  which have some overlap
the ideas in the present paper. Another piece of
related  which I became aware of during 
the final proof reading is \cite{howe} which extends
the ideas \cite{PS} on  the M-brane. In fact
on reduction to 4=1 dimensions
one gets a Born-Infeld action coupled to transverse scalars
and much of the analysis of the present paper goes through unchanged,
a fact also realized by Paulina Rychenkova and Miguel Costa.

Finally I would like to thank Dean Rasheed  for many useful
discussions and a great deal of 
help and assistance, particularly with checking
some detailed calculations. Some of the material in this paper will apper in his Ph.D. thesis.

\end{document}